\documentclass[12pt,a4paper]{article}
\usepackage{graphicx}
\usepackage{times}
\newcommand\aj{{AJ}}%
\newcommand\apj{{ApJ}}%
\newcommand\apjl{{ApJ}}%
\newcommand\aap{{A\&A}}%
\newcommand\mnras{{MNRAS}}%

\title{\bf Radio emission from the massive stars in Westerlund 1}
\author{S.M.~Dougherty$^{1,2}$, J.S.~Clark $^{3}$, I.~Negueruela$^4$, T.W.~Johnson$^{1,5}$ and J.M.~Chapman$^{6}$\\
\vspace{1cm}\\
\normalsize $^1$ NRC-HIA, Penticton, Canada \\
\normalsize $^2$ Institute for Space Imaging Science, University of Calgary, Canada\\
\normalsize $^3$ Open University, Milton Keynes, UK.\\
\normalsize $^4$ University de Alicante, Alicante, Spain \\
\normalsize $^5$ University of Victoria, Victoria, BC, Canada\\
\normalsize $^6$ Australia Telescope National Facility, Sydney, Australia}
\date{\mbox{}}
\begin{document}
\maketitle
\pagestyle{empty}
%
%
\def\bull{\vrule height .9ex width .8ex depth -.1ex}
\makeatletter
\def\ps@plain{\let\@mkboth\gobbletwo
\def\@oddhead{}\def\@oddfoot{\hfil\tiny\bull\quad
``The multi-wavelength view of hot, massive stars''; 39$^{\rm th}$ Li\`ege Int.\ Astroph.\ Coll., 12-16 July 2010 \quad\bull}%
\def\@evenhead{}\let\@evenfoot\@oddfoot}
\makeatother
%
%
\def\beginrefer{\section*{References}%
\begin{quotation}\mbox{}\par}
\def\refer#1\par{{\setlength{\parindent}{-\leftmargin}\indent#1\par}}
\def\endrefer{\end{quotation}}
%
%
{\noindent\small{\bf Abstract:} The diverse massive stellar
population in the young massive cluster Westerlund 1 (Wd~1) provides
an ideal laboratory to observe and constrain mass-loss processes
throughout the transitional phase of massive star evolution. A set of
high sensitivity radio observations of Wd~1 leads to the detection of 18
cluster members, a sample dominated by cool hypergiants, but with
detections among hotter OB supergiants and WR stars. Here the diverse
radio properties of the detected sample are briefly described. The
mass-loss rates of the detected objects are surprisingly similar
across the whole transitional phase of massive star evolution, at
$\sim10^{-5}$~M$_\odot$\,yr$^{-1}$. Such as rate is insufficient to
strip away the H-rich mantle in a massive star lifetime, unless the
stars go through a period of enhanced mass-loss. The radio luminous
star W9 provides an example of such an object, with evidence for two
eras of mass-loss with rates of $\sim10^{-4}$~M$_\odot$\,yr$^{-1}$.  }
%
%
\section{Introduction}

Mass-loss rate is a fundamental property of stars with masses in
excess of $\sim30~M_\odot$, important to stellar evolution and
feedback to the ISM. Mass-loss rates of massive main sequence stars
are insufficient to remove H-rich mantles to produce WR-type
stars. There must be epochs of enhanced mass-loss rate in their
evolution to WR stars, through phases that include hot supergiant
B[e], Luminous Blue Variables (LBV), cool Yellow Hypergiant (YHG) and
Red Supergiant (RSG) stars.

Westerlund 1 (Wd~1) is one of the most massive clusters in the Milky
Way ($M_{\rm total}\sim 10^5~M_\odot$), directly comparable to
Super-Star Clusters in other galaxies such as M82. Hence Wd1
represents a nearby example of one of these clusters, providing a
valuable opportunity to study the properties, evolution and
interaction of massive stars.

Wd~1 contains a unique population of cool and hot supergiants, with a
large population of post-MS stars that represent all phases of massive
star evolution: OB supergiants and hypergiants, RSGs, YHGs, and WR
stars (Clark \& Negueruela 2002; Clark et al. 2005).

Radio observations are a long established tool for estimating
mass-loss rates. Clark et al (1998) discovered two unusually radio
luminous stars in Wd~1 - the supergiant B[e] star W9 and the RSG
W26. Motivated by the possibility of detecting emission from stars
across a broad range of evolutionary stages, more sensitive radio
observations of Wd~1 were obtained with the Australian Telescope
Compact Array to establish radio characteristics of cluster members,
particularly their mass-loss rates. These observations have been used
in conjunction with optical, IR and X-ray observations from the
literature to elucidate the nature of the radio sources.

\begin{figure}[!ht]
\centering
\includegraphics[width=0.95\textwidth]{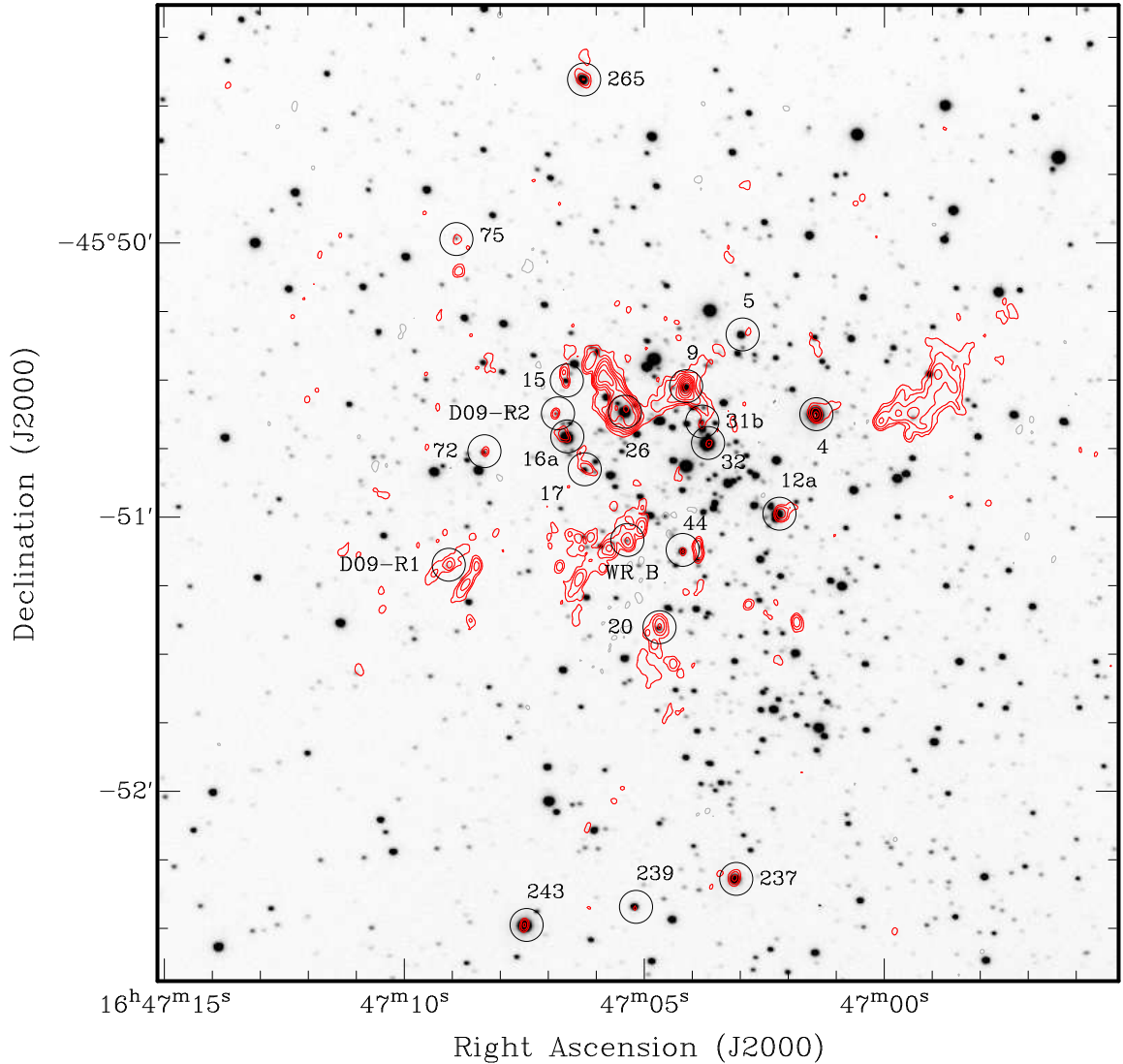}
\caption[]{8.6-GHz of Wd~1 overlaid on a FORS R-band image, with
 limiting magnitude ~17.5 mag. The radio sources with putative optical
 counterparts are identified by circles and Westerlund numbers}
\end{figure}

\section{Results}
\begin{itemize}

\item 18 cluster members were detected, dominated by cool hypergiants,
though with detections among hotter OB supergiants and WR stars (see Table
1).

\item The sources are a diverse population of point-like, unresolved
sources and extended, resolved sources with spectral indices
corresponding to thermal, non-thermal and composite thermal and
non-thermal emission.

\item The radio observations in conjunction with X-ray and IR
observations provide striking evidence for a high fraction of binaries
among the massive stars in Wd~1, strongly supporting previous
estimates of high binary fractions, in excess of 40\% (Clark et al,
2008; Ritchie et al. 2009).

\item The derived mass-loss rates determined for YHG W4, LBV W243 and
WR~L, which form an evolutionary sequence in some schemes, are all
closely the same.

\item Mass-loss rates of $\leq$10$^{-5}$M$_{\odot}$yr$^{-1}$ are
likely insufficient to remove the H-rich mantle unless stars remain in
the transitional phase for significantly longer than expected. This
implies an additional mechanism is required to shed the requisite
mass, with short-lived episodes of greatly enhanced mass loss.  Indeed
mass-loss rates $>10^{-4}$M$_{\odot}$yr$^{-1}$ have already been
inferred for RSGs e.g. VY CMa (Smith et al. 2001) and directly
observed for the YHG $\rho$ Cas (Lobel et al. 2003). The nebulae
around the RSGs W20, 26 and 236 already indicate that significant mass
loss has occurred for some stars within Wd~1, while the mass-loss rate
inferred for W9, over a magnitude greater than any other transitional
star in Wd~1, is of obvious interest.

\end{itemize}
\begin{table}
\centering
\caption[]{Number of radio emitters of given spectral type}
\begin{tabular}{lccc}
\hline
Spec.  & No. radio & Source &Cluster \\
Type   & emitters  & ID     &total \\
\hline
OB SGs & 3(4?) & 15, 17, D09-R1, (D09-R2?) & $\sim$150  \\
sgB[e] & 1 & 9 & 1 \\
BHGs   & 1 & 243 & 4 \\
YHGs   & 4(5?) & 4, 12a, 32, 265 (16a?) & 6 \\
RSGs   & 4 & 20, 26, 75, 237 & 4 \\
WN9-10 & 1 (2?) & 44, (5?) & 2 \\
WN5-8  & 3 & WR~B, 31b, 72 &14 \\
WC     & 1 & 239 & 8 \\
\hline
\end{tabular}
\end{table}

\subsection{W9 - a luminous radio source}

W9 is the brightest radio source in Wd~1, with a total flux at 8.6 GHz
of 55.4~mJy, which implies a luminosity of
$1.6\times10^{21}$~erg~s$^{-1}$. This makes W9 one of the most luminous
radio stars, being a factor of a few less luminous than the extreme
LBV $\eta$~Car at radio minimum (Duncan \& White 2002).

The radio structure of W9 is a compact radio source surrounded by an
extended emission region. The spectral index of the compact source
($+0.68\pm0.08$) is consistent with thermal emission from a stellar
wind.  The extended region is essentially flat ($+0.16\pm0.07$) and
arguably consistent with optically-thin thermal emission. Assuming a
radial ion distribution in the extended region goes as $r^{-2}$, the
lack of a turnover in its continuum spectrum implies that the region
is optically-thin down to at least 1.4 GHz, with an inner radius to
the region larger than the $\tau_\nu=1$ surface at 1.4 GHz. We believe
this is from an earlier epoch of mass loss, prior to the start of the
current stellar wind phase.

Modelling the envelope as a stellar wind surrounded by an optically
thin shell-like wind from an earlier phase of mass loss gives
mass-loss rates of $9.2\pm0.4 \times 10^{-5}(v_\infty/200~{\rm
km\,s}^{-1})f^{1/2}$ and $33\pm10 \times 10^{-5}(v_\infty/200~{\rm
km\,s}^{-1})f^{1/2}$~M$_\odot$~yr$^{-1}$ respectively, where $f$ is
the volume filling factor (see Dougherty et al. 2010 for modelling
details).

The structure of the envelope is similar to some LBVs, with W9 having
a current mass-loss rate similar to galactic examples e.g. Pistol
star. The rate deduced for the extended shell is close to the limit
expected for line-driven winds (Smith \& Owocki, 2006), and comparable
to several other galactic LBV's during outburst, though orders of
magnitude less than for P Cygni and $\eta$~Car during outburst (Clark
et al. 2009). Nevertheless, the question of whether W9 is undergoing
an `eruptive' event remains.

The X-ray properties imply W9 is a binary system, being too hard ($kT
\sim 3$~keV) and bright ($L_x\sim 10^{33}$~erg\,s$^{-1}$) to come from
a single star. However, it is not possible to constrain the properties
of a putative companion star from current observations. Given the
unusually high radio luminosity and concomitant high mass-loss rates,
W9 is clearly an object for further study.

\begin{figure}[!ht]
\centering
\includegraphics[width=0.6\textwidth,clip]{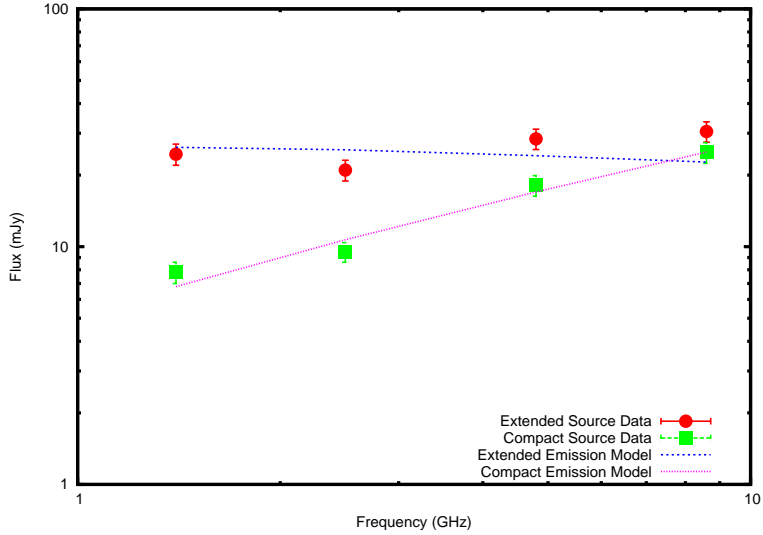}
\caption[]{Model fits to the observations of W9, using a partially
optically-thick compact source from a current stellar wind, and
a flat spectrum extended region due to optically-thin emission from 
an earlier epoch of mass loss via a stellar wind.}
\end{figure}

\subsection{OB stars}
Surprisingly, three of the $>100$ evolved OB stars are detected, with
radio fluxes larger than an order of magnitude higher than expected
for stellar winds in these types. These may be colliding-wind binaries
based on potentially composite spectra (both thermal and non-thermal
components) but there are neither X-ray emission nor RV variations
associated with these three objects to support the
claim. Alternatively, the high level of radio emission maybe
influenced by extended radio emission in which they are embedded.

\subsection{Red Supergiants}
All four RSGs are detected, three of them associated with large
nebulae with masses up to $\sim0.3~M_\odot$. The nebulae
around W20 and W26 have a pronounced cometary morphology, suggesting
significant interaction with either the intracluster medium or cluster
wind. 

W237 shows less evidence for such interaction and has
a kinematic age of $\sim$3,600~yr and a {\em time averaged} mass-loss
rate of $2\times10^{-5}({\rm v}_{\infty}/30{\rm
km\,s}^{-1})f^{1/2}$~M$_{\odot}$yr$^{-1}$. This is consistent with other
field RSGs, although it is substantially lower than inferred for
NML~Cyg and VY~CMa during the formation of their nebulae.

\subsection{Yellow Hypergiants}

The YHG W4 has a stellar wind with a mass-loss rate of
$10^{-5}($v$_{\infty}/200{\rm
km\,s}^{-1})f^{1/2}$~M$_{\odot}$yr$^{-1}$, consistent with the few
estimates available for other field YHGs. The extended nebulae
associated with W4, W12a and W265 are significantly less massive than
those associated with the RSGs in Wd~1, and likely arise from
quiescent mass loss rather than during outburst episodes.

Neither the YHGs nor RSGs are hot enough to ionize their
own stellar winds and/or more extended nebulae, and the requisite ionizing
photons must arise from either an unseen companion or the cluster
radiation field.

\subsection{B Hypergiants}
Of the extreme B-type hypergiants, only LBV W243 was detected, with a
spectral index consistent with thermal emission.  The corresponding
mass-loss rate is comparable to YHG W4, as expected given the
similarity in current spectral type and radio flux. Upper limits for
the three other B hypergiants were found to be
$2\times10^{-6}($v$_{\infty}/200{\rm
km\,s}^{-1})f^{1/2}$~M$_{\odot}$yr$^{-1}$ , consistent with mass-loss
rates amongst field stars of these types

\subsection{Wolf-Rayets}
Five of the 24 WRs in Wd~1 were detected. WR L has a partially
optically-thick wind, with a mass-loss rate consistent with stars of
identical spectral type in the Galactic Centre cluster and the general
field population ie. $\dot M =2\times 10^{-5}({\rm v}_{\infty}/1000{\rm
km\,s}^{-1}) f^{1/2}$~M$_{\odot}$yr$^{-1}$ .  The remaining three
(WR~A, B and V) are identified as having composite spectra from a
CWB. The optical and X-ray properties of WR A and WR B have previously
indicated these to be binaries, while this is the first hint of
binarity in WR V.
%
%
%
\section*{Acknowledgments}
We thank Paul Crowther, Ben Davies, Simon Goodwin, Rene
Oudmaijer, Julian Pittard, Ben Ritchie and Rens Waters for many
stimulating discussions related to this work.
A special thanks to Rob Reid for advice on smear fitting, the deconvolution method used.
 The Australia Telescope Compact Array is part of the
Australia Telescope which is funded by the Commonwealth of Australia
for operation as a National Facility managed by CSIRO. This work was
partially supported by a UK Research Council (RCUK) Fellowship and by
the Spanish Ministerio de Ciencia e Innovaci\'on (MICINN)
under grants AYA2008-06166-C03-03 and Consolider-GTC CSD2006-70.

%
\footnotesize
\beginrefer
\refer {Clark}, J.~S., {Crowther}, P.~A., {Larionov}, V.~M., {et~al.} 2009, \aap, 507,
  1555

\refer {Clark}, J.~S. \& {Negueruela}, I. 2002, \aap, 396, L25

\refer {Clark}, J.~S., {Fender}, R.~P., {Waters}, L.~B.~F.~M., {et~al.} 1998, \mnras,
  299, L43

\refer {Clark}, J.~S., {Negueruela}, I., {Crowther}, P.~A., \& {Goodwin}, S.~P. 2005,
  \aap, 434, 949

\refer {Clark}, J.~S., {Muno}, M.~P., {Negueruela}, I., {et~al.} 2008, \aap, 477, 147

\refer {Duncan}, R.~A. \& {White}, S.~M. 2002, \mnras, 330, 63

\refer {Lobel}, A., {Dupree}, A.~K., {Stefanik}, R.~P., {et~al.} 2003, \apj, 583, 923

\refer {Ritchie}, B.~W., {Clark}, J.~S., {Negueruela}, I., \& {Crowther}, P.~A.
  2009, \aap, 507, 1585

\refer {Smith}, N., {Humphreys}, R.~M., {Davidson}, K., {et~al.} 2001, \aj, 121, 1111

\refer {Smith}, N. \& {Owocki}, S.~P. 2006, \apjl, 645, L45

\endrefer           
\end{document}